\documentclass[pdflatex,twocolumn,sn-aps]{sn-jnl}

\usepackage{graphicx}%
\usepackage{multirow}%
\usepackage{amsmath,amssymb,amsfonts}%
\usepackage{amsthm}%
\usepackage{mathrsfs}%
\usepackage[title]{appendix}%
\usepackage{xcolor}%
\usepackage{textcomp}%
\usepackage{manyfoot}%
\usepackage{booktabs}%
\usepackage{algorithm}%
\usepackage{algorithmicx}%
\usepackage{algpseudocode}%
\usepackage{listings}%
\usepackage{natbib}
\usepackage{hyperref}

\begin{document}

\title[Article Title]{Size optimization for observeing Majorana fermions}


\author[1]{\fnm{Guo-Jian} \sur{Qiao}}\email{qiaoguojian19@gscaep.ac.cn}

\author[1]{\fnm{Zhi-Lei} \sur{Zhang}}\email{zhangzhilei21@gscaep.ac.cn}

\author[2]{\fnm{Xin} \sur{Yue}}\email{yuexin@csrc.ac.cn}

\author*[1]{\fnm{C. P.} \sur{Sun}}\email{suncp@gscaep.ac.cn}

\affil*[1]{\orgname{Graduate School of China Academy of Engineering Physics}, \orgaddress{\city{Beijing}, \postcode{100193}, \country{China}}}

\affil[2]{\orgname{Beijing Computational Science Research Center}, \orgaddress{\street{Street}, \city{City}, \postcode{10587}, \state{State}, \country{China}}}


\abstract{Majorana fermions (zero modes) are predicted to emerge in nanowire-superconductor heterostructures. This theoretical prediction typically relies on an oversimplified model, where both the nanowire and the superconductor are idealized as one-dimensional systems. In reality, heterostructures have finite sizes that deviate from this idealization—and as a result, smoking-gun evidence confirming the existence of these zero modes remains elusive. Here, we investigate the finite-size effects of both the nanowire and the superconductor, and optimize their sizes to ensure that only one Majorana fermion exists at each end of the heterostructure. It is discovered that the optimal transverse sizes of the nanowire are less than 100\,nm in width and approximately 1\,nm in thickness. For the superconductor layer, its optimal thickness (a key aspect of its size) must exceed its coherence length. We also present the optimal sizes of the two types of materials used in the experiment in a quantitative manner. Notably, the identified optimal thickness of the superconductor (Al films, $\sim$1000\,nm)—a critical size parameter—is two orders of magnitude larger than the thickness of Al films currently utilized in experimental devices (e.g., InSb-Al and InAs-Al heterostructures). Our findings could explain why Majorana fermions have not been observed in current experiments, and offer guidance for the size selection of heterostructures to implement Majorana fermions in future studies.}


\maketitle

\section*{Introduction}\label{sec1}
The experimental observation of Majorana fermion (zero mode) in nanowire-superconductor heterostructures has garnered sustained attention \cite{Mourik_2012,Deng_2012,Das2012,Nichele_2017,Gazibegovic2017,Zhang2018,Marco_2021_Nontopological_zero-bias,Zhang_2022,Vaiti_2020,Levajac2023,Dvir2023,Aghaee2025,zhang2025robustzeromodespbtepb}, primarily due to their non-Abelian statistics and potential applications in topological quantum computing \cite{A_Yu_Kitaev_2001,Kitaev_2003,Nayak_2008}. A minimal model has been well established for this type of heterostructure, consisting of a nanowire proximity-coupled to an \textit{s}-wave superconductor (SC) \citep{Oreg_2010,Alicea_2012,Lutchyn2018,Prada2020,Qiao_2022,Qiao_2024}. Notably, the nanowire is typically modeled as a one-dimensional system, while the SC is idealized either as a one-dimensional system  \cite{Qiao_2024} or as an infinitely large bulk \cite{Alicea_2012} in the minimal model \cite{Prada2020}. Based on this model, it has been shown that there exists one Majorana fermion (MF) localized at each end of the heterostructure. This MF is expected to be experimentally detected by the zero-bias peak (ZBP) with a height of $2e^{2}/h$ in very low temperature \cite{Li_2014,Qiao_2022,Zhang_Fan_2023,Yue_2023,Qiao_2024}. Consequently, the emergence of $2e^{2}/h$ ZBP and its plateau within specific parameter regions serve as a critical signature of observed MFs. However, the models used for the nanowire and SC are over-simplified. In practical heterostructures, both the nanowire and the SC have finite sizes. Their transverse sizes are $\sim$100\,nm and $\sim$10\,nm repectively, and their longitudinal size is $\sim$1$\mu$m \cite{Potter_2011,Lutchyn_2011_in_Multiband_Nanowires,Stanescu_2011_MF_in_semiconductor_nanowires,Liu_Jie_2012,Karsten_2018,Antipov_2018,Levajac2023}. These finite-size constraints in the transverse direction result in multiple subbands in both the nanowire and SC. The existence of multiple subbands can result in multiple MFs (or quasi-MFs \cite{Wimmer_2019,Prada2020,Zhang_2022}) localized at one end  \cite{Potter_2010,Roy_2013}. In such scenarios, the ZBP will exceed the quantized value of $2e^2/h$, which deviates from the theoretical predictions by the minimal model. Therefore, we need to study the finite-size effects of the nanowire and SC for emergence of MF, and optimize their sizes to ensure that only one Majorana fermion exists at each end of the heterostructure. 

In recent years, numerous studies on finite-size effects of nanowire \cite{Lutchyn_2011_in_Multiband_Nanowires,Stanescu_2011_MF_in_semiconductor_nanowires,Potter_2011,Liu_Jie_2012,Roy_2013} and even SC \cite{Jelena_2017_Finite_size_effects,Reeg_2018_Metallization_nw,Legg_2022_Metallization_Ti_NW} have been conducted, and they primarily focus on the energy bands of the heterostructure and their defined boundaries of the topological phases \citep{Karsten_2018,Antipov_2018,Winkler_2019}. However, these studies do not explicitly determine, in principle, the number of MFs that exist in the topological phase. Since the strengths of experimental signals (e.g., the height of ZBP) depend on the number of MFs, it remains unclear whether the observed signatures in current finite-size heterostructures can be attributed to the presence of MFs. Therefore, determining the number of MFs in both finite-size nanowires and SCs is critical for clarifying the interpretation of current experimental signals.

Here, we not only analytically define the boundary of the topological phase for finite-size heterostructures, but also determine the number of localized MFs in various phase regions by using proposed method for defining dressed MFs \cite{Qiao_2024,ZZ2025,yue2025}. We also analytically obtained the optimized transverse size of the nanowire to realize only one MF at each end of the heterostructure. The optimized transverse sizes are such that the width is less than the critical size (approximately 100\,nm), and the thickness is nearly zero (around 1\,nm).

Additionally, we predict that as the size of SC increases, the oscillation of the shift in the chemical potential and that of the effective gap will decay to a steady value. The period of these oscillations corresponds to half the Fermi wavelength, while the characteristic decay length is the superconducting coherence length. Both the oscillation amplitude and the steady value depend on the proximity coupling strength. In reality, it is not feasible to control the superconducting thickness with a precision of 1 $\text{\AA}$, which is significantly smaller than the Fermi wavelength ($\sim$ 1nm) \cite{Xue2004}. Therefore, the optimal superconducting size (its thickness) should be larger than the superconducting coherence length. Beyond this thickness, the shift in chemical potential and the effective energy gap will stabilize, and their magnitude can be tuned via the strength of proximity coupling. 

Since above these findings are entirely determined by the intrinsic properties of SC and strength of proximity coupling, while being independent of the nanowire. As a result, they are generally valid for other superconducting heterostructures, including insulator-superconductor heterostructures \cite{Yue_2023,yue2025} and systems with two quantum dots coupled to a superconductor (two QD-SC systems) \cite{Dvir2023,zhang2025robustzeromodespbtepb}. A summary of the optimal sizes for the two types of heterostructures is provided in Tab. \ref{tab1}. Notably, for the heterostructures of InAs-Al and InSb-Al, the optimized thickness ($\sim$1000nm) is two orders of magnitude larger than that of the currently used Al films ($\sim$10nm). This insufficient thickness could explain why smoking gun evidence of MFs has not yet been observed in these systems. Specifically, at a thickness of approximately 10\,nm, the significant shift in the chemical potential and the excessively large effective gap in the nanowire both suppress the emergence of Majorana fermions.

\section*{Results}\label{sec2}
\subsection*{Topological phase vs number of MFs}
\begin{figure}
\includegraphics[scale=1.0]{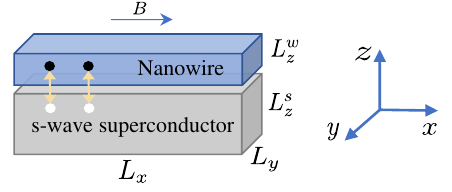}
\caption{A finite-size nanowire is in contact with a finite-size $s$-wave superconductor. The sizes in the $y$ and $z$ directions are finite and significantly smaller than the size in the $x$ direction. The sizes in the $x$ and $y$ direction are identical for both the nanowire and the superconductor, i.e., $L_{x}^{w}=L_{x}^{s}$ and $L_{y}^{w}=L_{y}^{s}$.}	
\label{fig:finite_size_Nw_and_SC}
\end{figure}

We characterize the nanowire-superconductor heterostructure by the Slab model \cite{Antipov_2018,Karsten_2018}, where both the nanowire and SC have finite sizes, as shown Fig. \ref{fig:finite_size_Nw_and_SC}. The Hamiltonian of this heterostructure in lattice space is
\begin{align}
H=H_{w}+H_{s}+H_{t}.
\label{eq:Hybrid}
\end{align}
Here, $H_{\alpha}$ with $\alpha(=w,s)$ respectively represents the Hamiltonians of the nanowire and SC, and $H_{t}$ describes the proximity tunneling between them. More details about the model can be found in the Method Section. Due to finite-size constraints in $y-z$ direction, discrete modes of electrons propagating transversely are generated \citep{Potter_2010,Lutchyn_2011_in_Multiband_Nanowires,Stanescu_2011_MF_in_semiconductor_nanowires}. The transverse momentum of these modes is $\boldsymbol{k}_{\perp}^{\alpha}=(k_{y}^{\alpha},k_{z}^{\alpha})$, where  $k_{i}^{\alpha}=\pi l_{i}^{\alpha}/(N_{i}^{\alpha}+1)$ corresponds to the $l_{i}^{\alpha}$-th discrete momentum mode. Here, $l_{i}^{\alpha}=1,2,\ldots,N_{i}^{\alpha}$, and the notation $N_{i}^{\alpha}$ with $i\in\{x,y,z\}$
represents the total number of sites along the $i$-direction. 

Then, we demonstrate how these transverse modes affect the energy band of the nanowire and SC. Under periodic boundary conditions along the $x$-direction,  the kinetic energy of electrons in nanowire or SC is $\varepsilon_{\alpha}(\boldsymbol{k}_{\alpha}) =\epsilon_{\alpha}(k_{x})+\epsilon(\boldsymbol{k}_{\perp}^{\alpha})-\mu_{\alpha},\label{eq:electron_Kinetic}$
where $\epsilon_{\alpha}(k_{x})$ and $\epsilon(\boldsymbol{k}_{\perp}^{\alpha})$  represent the electron kinetic energies along the $x$-direction and transverse directions, respectively.  Here, $\boldsymbol{k}_{\alpha}=(k_{x},\boldsymbol{k}_{\perp}^{\alpha})$, where $k_{x}$ is the electron's momentum component in the $x$-direction, and $\mu_{\alpha}$ is the chemical potential. By diagonalizing the Hamiltonian of the nanowire, its spectrum is given by  
\begin{equation}
E_{\pm}(\boldsymbol{k}_{w})\simeq \varepsilon_{w}(\boldsymbol{k}_{w})\pm\sqrt{h_{w}^{2}+(\alpha_{x}\sin k_{x})^{2}}. 
\label{Nw_spectrum} 
\end{equation}
Here, $h_{w}$ is the Zeeman energy of nanowire, $\alpha_{x}$ is Rashba spin-orbit coupling in the $x$ direction, and the spin-orbit coupling in the $y$-direction is neglected for simplicity. It is seen from Eq. \eqref{Nw_spectrum} that, for a given set of transverse momentum modes $\boldsymbol{k}_{\perp}^{\alpha}$, $E_{\pm}(k_{x},\boldsymbol{k}_{\perp}^{\alpha})$ forms a sub-band as $k_{x}$ varies from $-\pi$ to $\pi$. When the kinetic energies of two transverse subbands are equal or close, the subbands in the nanowire become degenerate or near degenerate. Similarly, these phenomena manifest in SC as a result of finite-size effects. How do the multiple subbands originating from finite-size effects influence the topological properties of the entire heterostructure? 

For band topology, it can be characterized by topological invariant \citep{Tewari_2012,Budich_2013,Chiu_2016_Classification_topological_symmetries}. By the defined topological invariant, its analytical result is shown as (see Supplemental Information):
\begin{equation}
M=\prod_{\boldsymbol{k}_{\perp}^{w}}\mathrm{sgn}(\prod_{k_{x}=0,\pi}[h_{\mathrm{eff}}^{2}(\boldsymbol{k}_{w})-\varepsilon_{\mathrm{eff}}^{2}(\boldsymbol{k}_{w})-\Delta_{\mathrm{eff}}^{2}(\boldsymbol{k}_{w})]),\label{eq:TIn}
\end{equation}
In Eq. (\ref{eq:TIn}), the effective Zeeman energy and pairing strength are $h_{\mathrm{eff}}=h_{w}+\lambda(\boldsymbol{k}_{w})h_{s}$ and $\Delta_{\mathrm{eff}}=\lambda(\boldsymbol{k}_{w})\Delta_{s}$, where $h_{\alpha}$ is the Zeeman energy \cite{Clogston_1962,Maki1964,Reeg_transport_Signature_2017,Marcus_Effective_g_2018}, and $\Delta_{s}$ is the superconducting pairing strength. And the effective kinetic energy is given by 
$\varepsilon_{\mathrm{eff}}(\boldsymbol{k}_{w})=\varepsilon_{w}(\boldsymbol{k}_{w})-\mu_{\mathrm{shift}}(\boldsymbol{k}_{w})$, where the shift in the chemical potential of nanowire are caused by the all superconducting subbands:  
\begin{equation}
	\mu_{\mathrm{shift}}(\boldsymbol{k}_{w})=\sum_{k_{z}^{s}}\frac{|T_{k_{z}^{w},k_{z}^{s}}|^{2}}{E_{s,+}(\boldsymbol{k}_{s})E_{s,-}(\boldsymbol{k}_{s})}\epsilon_{s}(\boldsymbol{k}_{s}).\label{eq:effective_kinetic}
\end{equation}
Here, $E_{\pm}^{s}(\boldsymbol{k}_{s})=\sqrt{\varepsilon_{s}^{2}(\boldsymbol{k}_{s})+\Delta_{s}^{2}}\pm h_{s}$
is the quasi-excitation energy in SC, and $T_{k_{z}^{w},k_{z}^{s}}$ represents the tunneling strength between different subbands of the nanowire and SC. 
In the derivation above, we have assumed that the tunneling processes involving different momenta in the $z$-direction between the nanowire and SC do not have any interferences or correlations with each other,
i.e., $T_{k_{z}^{w},k_{z}^{s}}T_{\tilde{k}_{z}^{w},\tilde{k}_{z}^{s}}^{*}\simeq\delta_{k_{z}^{w}\tilde{k}_{z}^{w}}\delta_{k_{z}^{s}\tilde{k}_{z}^{s}}|T_{k_{z}^{w},k_{z}^{s}}|^{2}$
\citep{Sau_Robustness_MF_2010,Qiao_2022}. Meanwhile, the correction
factor emerging in the above formula is defined as $\lambda(\boldsymbol{k}_{w})=\sum_{k_{z}^{s}}|T_{k_{z}^{w},k_{z}^{s}}|^{2}/E_{s,+}E_{s,-}$. This factor reveals that the dressed effect of SC \cite{Qiao_2024} on the Zeeman energy
($\lambda(\boldsymbol{k}_{w})h_{s}$) and pairing strength ($\lambda(\boldsymbol{k}_{w})\Delta_{s}$)
in the nanowire depends on all superconducting sub-bands. Notably,
the primary contributions of the dressed effect on the nanowire originate
from the subbands near the superconducting Fermi surface. Since the
correction factor depends on the specific subband of the nanowire, identified
by $\boldsymbol{k}_{w}^{\perp}$, the effective Zeeman energy $h_{\mathrm{eff}}(\boldsymbol{k}_{w})$
and pairing strength $\Delta_{\mathrm{eff}}(\boldsymbol{k}_{w})$
exhibit variations across different subbands. This discovery provides
an analytical corroboration of previously reported numerical results
\cite{Antipov_2018}. 

Based on the obtained topological invariant \eqref{eq:TIn}, $M=-1$ defines the parameter region corresponding to the topological phase,
whereas $M=1$ defines the parameter space of the non-topological phase. The parameter space \cite{Parameterspace} refers to the chemical
potential and Zeeman energy in nanowire and SC, represented by $(\mu_{w},h_{w},\mu_{s},h_{s})$. This topological phase diagram is valid for any finite-sized nanowire-superconductor heterostructure, even in cases where the proximity-tunneling strength between the nanowire and SC is strong. It is evident that the above topological phase does not determine the number of MFs. Since the number of MFs directly determines their observable signal in experiments, then we need to determine the number of MFs in this topological phase.

Following the method proposed in Ref. \cite{Qiao_2024}, we define dressed MFs by $\eta_{\nu}^{\dagger}=\eta_{\nu}$, where $\eta_{\nu}$ is the annihilation operator of the quasiparticle of the total hybrid system. By this definition, we obtain that, for any subband of the
nanowire denoted by the transverse mode $\boldsymbol{k}_{w}^{\perp}$, the condition (see Supplementary Information)
\begin{equation}
	\prod_{k_{x}=0,\pi}\mathrm{sgn}[h_{\mathrm{eff}}^{2}(\boldsymbol{k}_{w})-\varepsilon_{\mathrm{eff}}^{2}(\boldsymbol{k}_{w})-\Delta_{\mathrm{eff}}^{2}(\boldsymbol{k}_{w})]<0\label{eq:existed_condition}
\end{equation}
determines the parameter region in the space of $(\mu_{w},h_{w},\mu_{s},h_{s})$. In this region, there exist one MF localized at each end of nanowire
and SC. Naturally, if the chemical potential and magnetic field in nanowire and SC are adjusted such that there exist $\mathcal{N}$
transverse mode with different $k_{w}^{z}$ satisfying Eq. \eqref{eq:existed_condition},
then there will consequently exist $\mathcal{N}$ MFs localized at the each end. The $\mathcal{N}$ MFs can give rise to a ZBP, with the height
of $\mathcal{N}(2e^{2}/h)$ at zero temperature \citep{Wimmer_2011,Roy_2013}.
Furthermore, it is easy to see that Eq. \eqref{eq:existed_condition}
is precisely each factor in the consecutive product of subbands in
the expression for the topological invariant \eqref{eq:TIn}. Consequently,
when the odd (even) subbands with different $k_{w}^{z}$ satisfy Eq. \eqref{eq:existed_condition},
the system is in the topological (non-topological) phase. From the
above results, we conclude the relationship between the number of
MFs, topological invariants, and the height of the ZBP in a finite-size heterostructure as follows:
\begin{enumerate}
	\item The presence of an odd (even) number of MFs localized at each end
	corresponds to a topological (non-topological) phase.
	\item In the topological (non-topological) phase, there exist an odd (even) number
	of MFs are localized at each end. However, the specific number of
	MFs is not determined.
	\item The $\mathcal{N}$ MFs give rise to a ZBP with a height $\mathcal{N}(2e^{2}/h)$,
	where $\mathcal{N}$ is an odd (even) integer. 
\end{enumerate}
The number of MFs can predict the experimentally observed signal (such as the height of ZBP), whereas the topological phase
diagram cannot. Hence, our developed theory for determining the number of MFs within the parameter space serves as the foundation for their experimental observation.

It is worth emphasizing that the analytical result for determining number of MFs {[}see Eq. \eqref{eq:existed_condition}{]} are valid to finite-size heterostructures under arbitrary proximity tunneling strengths. Namely, when the chemical potential, magnetic field in nanowire and SC, along with the proximity tunneling strengths between them are specified, the number of MFs is directly determined by these parameters. In the 1D limit of the nanowire and SC, i.e., $N_{y}^{\alpha}=N_{z}^{\alpha}=1$ with $\alpha=w,s$, the finite-size effects are neglected. In this limit, Eq. \eqref{eq:existed_condition} simplifies to results previously reported in Ref. \cite{Qiao_2024}. In this scenario, only one MF
can exist at each end of the nanowire-superconductor heterostructure, resulting in
a ZBP with a quantized height of $2e^{2}/h$ \cite{Lutchyn_2011_in_Multiband_Nanowires,Qiao_2024}. When the finite-size effects are taken into account, they give rise to transverse subbands in both the nanowire and SC. These transverse subbands may increase the number of MFs at each end, thereby leading to distinct experimental signatures, such as ZBPs with a height
of $4e^{2}/h$ or even $6e^{2}/h$. Next, we show how to optimize the transverse size of nanowire and SC to  ensure that only one Majorana fermion exists at each end of the heterostructure, and further determine their optimal sizes for observing MFs.
\begin{table*}[!htbp]  
	\centering  
	\resizebox{\linewidth}{!}{  
		\begin{tabular}{c|c|c|c|c|c}  
			\hline  
			Heterostructure  & Platform&Thickness of SC ($L_{z}^{s}>\xi_s$) & Size of NW ($L_{y}^{w}<L_{c},L_{z}^{w}\rightarrow 0$) &MF Type&Observed Signal\\
			\hline
			\multirow{5}{*}{NW-SC}&InSb-NTiNb &$L_{z}^{s}>$ 5 nm & $L_{y}^{w}<$74 nm & \multirow{5}{*}{MZM} &\multirow{5}{*}{$\frac{2 e^2}{h} $ ZBP}\\
			&InSb-Nb &  $L_{z}^{s}>$180 nm & $L_{y}^{w}<$74 nm &  &\\
			&InSb-Al  & $L_{z}^{s}>$1600 nm & $L_{y}^{w}<$74 nm & & \\
			&InAs-Al  &  $L_{z}^{s}>$1600 nm & $L_{y}^{w}<$97 nm &  &\\
			\hline
			\multirow{2}{*}{TI-SC}&$\text{Bi}_2 \text{Te}_{3}$-$\text{Nb}\text{Se}_2$  & $L_{z}^{s}>$10 nm & \multirow{2}{*}{$\times$} & MVS &$2 e^2/h $ ZBP\\
			&$\mathrm{Cr:(Bi,Sb)_2 Te_3}$-$\text{Nb}$  & $L_{z}^{s}>$180 nm &  & CMF & $e^2/(2h)$ CP\\
			\hline
		\end{tabular}
	}  
	\caption{For rectangular geometry (see Fig. \ref{fig:finite_size_Nw_and_SC}), the optimal transverse sizes of nanowires (NWs) and superconductors (SCs) in various widely adopted heterostructures. In these sizes, a Majorana zero mode (MZM) is localized at each end when the system is in the topological phase, and its observed signal is a zero-bias peak (ZBP) with a height of $2e^2/h$. The optimal thickness of SC in topological insulators (TI) and SC heterostructures are given. This platform supports the emergence of chiral Majorana fermions (CMF) or Majorana vortex states (MVS), and their observed signals are half-integer conductance peaks (CP) or $2e^2 /h$ ZBP at the center of MVS, respectively.}  
	\label{tab1}  
\end{table*}
\subsection*{Size Optimization of Nanowire and SC}
The bandwidths of the nanowire and SC as the maximum energy scale are much larger than other parameters such as Zeeman energy, chemical potential, and the superconducting gap. Consequently, the corrections from the high-energy excitations in SC to the nanowire are small and the factor with $k_{x}=\pi$ in Eq. \eqref{eq:existed_condition} is always negetive \cite{Lutchyn_2011_in_Multiband_Nanowires,Qiao_2024}. Then the condition for determining the number of MFs simplifies to

\begin{equation}
	|\epsilon_{w}(\boldsymbol{k}_{w}^{\perp})-\mu_{w}+\mu_{\mathrm{shift}}(\boldsymbol{k}_{w}^{\perp})|<\mu_{c}(\boldsymbol{k}_{w}^{\perp})/2,
	\label{eq:existed_condition-1}
\end{equation}
where 
\begin{equation}
	\mu_{c}(\boldsymbol{k}_{w}^{\perp})=2\sqrt{[h_{w}+\lambda(\boldsymbol{k}_{w}^{\perp})h_{s}]^{2}-(\lambda(\boldsymbol{k}_{w}^{\perp})\Delta_{s})^{2}}\label{eq:chemical_window}
\end{equation}
determines the range (window) of chemical potential of existence of
MFs as $\mu_{c}$. Here, $\epsilon_{w}(\boldsymbol{k}_{w}^{\perp})$,
$\lambda(\boldsymbol{k}_{w}^{\perp})$ and $\mu_{\mathrm{shift}}(\boldsymbol{k}_{w}^{\perp})$
are obtained by setting $k_{x}=0$. While even
transverse subbands satisfy Eq. (\ref{eq:existed_condition-1}), the
system transitions into a non-topological phase, in which even localized MFs emerge at each end. To achieve only one Majorana fermion at each end, the transverse sizes of the nanowire need to be optimized to avoid extensive low-energy subband degeneracy ($l_{i}^{w}\ll N_{i}^{w}$) in nanowire. Thus, the transverse sizes of the nanowire must satisfy the following union of inequalities derived from $p$ ranging over all positive integers (see Supplementary Information):
\begin{equation}
L_{w}^{z}<L_{w}^{y}<L_{c},\quad|L_{w}^{y}-pL_{w}^{z}|>\frac{L_{w}^{y}}{6p^{3}}(\frac{L_{w}^{y}}{L_{c}})^{2}.
\label{eq:size_constrant}
\end{equation}
For the case where  $L_{w}^{y}<L_{w}^{z}<L_{c}$, the results can be obtained by swapping $L_{w}^{y}$ and $L_{w}^{z}$  in Eq. \eqref{eq:size_constrant}. Here, the critical length $L_{c}=\sqrt{\hbar^{2}\pi^{2}/2m_{w}\mu_{\mathrm{max}}}$ is determined by the maximum window of chemical potential $\mu_{\mathrm{max}}$. When $h_w h_s <0$, the maximum window of chemical potential is given by Eq. \eqref{eq:chemical_window} as $\mu_{\mathrm{max}}=2h_{w}(B_c)$, where $B_c$ is the critical magnetic field of SC. This indicates that a larger chemical potential window requires smaller transverse sizes. It follows from Eq. (\ref{eq:size_constrant}) that, the ratio of the sizes in the $y$ to $z$ directions should not be close to an integer. In this condition, the greater the difference between them, the better. From this, we can further deduce that the more symmetrical the transverse geometry of the nanowire is, such as in circular \cite{Das2012,Zhang2018,Vaiti_2020,Marco_2021_Nontopological_zero-bias,Zhang_2022} or hexagonal \cite{Levajac2023,Nick_2023}, the more likely subband degeneracy is to occur. Therefore, more symmetrical geometries are unfavorable for observing MFs. For rectangular geometries, the optimal transverse size of the nanowire are $L_{w}^{y}<L_{c}$ and $L_{w}^{z}\rightarrow 0$. 

Then we further optimize the superconducting size (i.e., its thickness) to observe MFs. Given that the main contributions of the dressed effect to the nanowire originate from the subbands near the superconducting Fermi surface, the tunneling strength between SC and the different subbands of the nanowire is considered homogeneous, i.e., $T_{k_{z}^{w},k_{z}^{s}}\simeq T_{k_{z}^{w}}/\sqrt{N_{s}^{z}+1}$ \cite{yue2025}. In this situation, the shift in the chemical potential and the effective gap can be analytically derived (see Supplemental Information). When $\mu_s \gg \Delta_s\gg h_s$ and $L_{s}^{z} \gtrsim \xi_s $, where $\xi_s=\hbar v_{F}/(2\Delta_{s})$ is the coherence length of SC, for different low-energy subbands with $l_{y}^{w}\ll N_{y}^{w}$, the shift in the chemical potential in the nanowire is simplified as  
\begin{equation}
	\mu_{\mathrm{shift}}\simeq 2\Delta\sin(2k_{F}L_{s}^{z})e^{-\frac{L_{s}^{z}}{\xi_{s}}},
	\label{mu_shift}
\end{equation}
and the effective gap is determined as 
\begin{equation}
\Delta_{\mathrm{eff}}\simeq \Delta [1+2\cos(2k_{F}L_{s}^{z})e^{-\frac{L_{s}^{z}}{\xi_s}}].
\label{induced_gap}
\end{equation}
Here, $\Delta=(a/\xi_{s})|T_{k_{z}^{w}}|^{2}/\Delta_{s}$ depends on the tunneling strength between the nanowire and SC, which characterizes the dressed effect of SC. 

It follows from Eqs. (\ref{mu_shift}, \ref{induced_gap}) that the oscillation of the chemical potential shift and that of the effective gap decay to a steady value as the SC thickness increases, where the characteristic length of the decay is the superconducting coherence length. When $L_{z}^{s}\lesssim \xi_{s}$, for the given subband of nanowire, the shift of chemical potential and effective gap oscillate rapidly with respect to $L_{z}^{s}$, and the period of the oscillation is half the Fermi wavelength $\lambda_{F}/2\sim 1\mathrm{nm}$ \cite{Jelena_2017_Finite_size_effects,Reeg_2018_Metallization_nw,yue2025}. The oscillatory peaks of the chemical potential shift and the effective gap occur near the resonance thicknesses $L_{z}^{s}=(n/2+1/4)\lambda_{F}$ and $L_{z}^{s}=n\lambda_{F}/2$, where $n$ is a positive integer. The amplitude of the oscillation and the magnitude of the steady value depend on the strength of the proximity coupling. In fact, tuning the thickness of superconducting thin films within half the Fermi wavelength ($\sim$1nm) is not feasible  \cite{Xue2004}. Thus, the optimized superconducting size should exceed the critical thickness of SC, $\xi_s$ \cite{Jelena_2017_Finite_size_effects}. At this thickness, the magnitude of the effective gap is further adjusted by the proximity coupling strength. Since these findings are entirely determined by the properties of SC and proximity coupling, and are independent of the nanowire, they are generally valid for other superconducting heterostructures, such as insulator-superconductor heterostructures \cite{Yue_2023,yue2025} and two QD-SC systems \cite{Dvir2023,zhang2025robustzeromodespbtepb}. 
\begin{figure*}
	\includegraphics[width=1.0\textwidth]{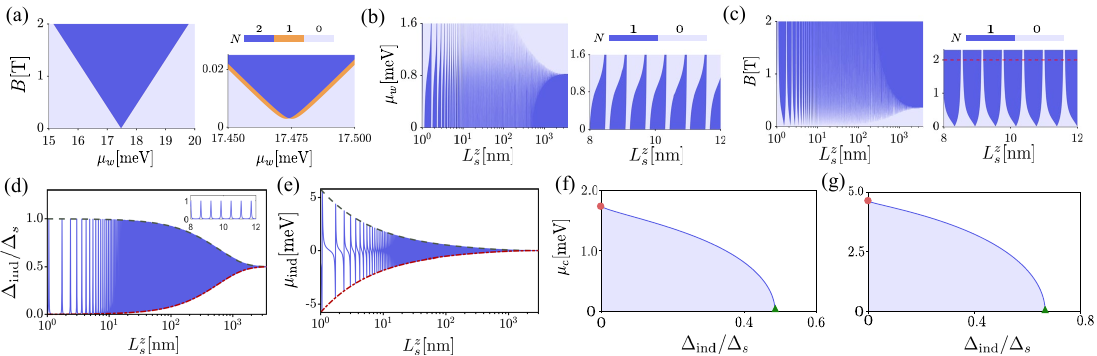}
	\caption{(a) For an InSb nanowire with $L_{w}^{y}=L_{w}^{z}=70\mathrm{nm}$, coated with a 10\,nm thick Al film, the number of Majorana fermions in the parameter of $\mu_w$ and $B$. (b) For an InSb nanowire with $L_{w}^{y}=70$\,nm and $L_{w}^{z}=1$\,nm, the number of Majorana fermions varies with changes in superconducting thickness and chemical potential in the nanowire at a constant magnetic field $B=0.8$T. (c) For a given chemical potential of the nanowire $\mu_w=0.5\Delta_s$, the number of Majorana fermions is determined by the superconducting thickness and the applied magnetic field. For a given subband of nanowire characterized by $\boldsymbol{k}_{w}=(2\pi/(N_{y}^{w}+1),k_{z}^{w})$, the induced chemical potential (d) and the induced gap (e) vs. thickness of superconductor. (f) and (g) display the constraint relation for the chemical potential window and the induced energy gap for InAs-Al and InSb-Al. The maximum of chemical potential window reaches 1.73(4.63) meV for InAs-Al (InSb-Al), as indicated by the red circular dots. The maximum of the induced energy gap is 0.48$\Delta_{s}$ (0.67$\Delta_{s}$) for InAs-Al (InSb-Al), indicated by green triangles. The parameters are set as $\Delta_{s}=0.34\,\mathrm{meV}$, $\mu_{s}=1\,\mathrm{eV}$, $a=0.85\mathrm{\AA}$, $t_s \simeq 0.015t_w=10\,\mathrm{eV}$ and $T_{k_{z}^{w}}\simeq 27.5\,\mathrm{meV}$.}
	\label{fig:Finite_size_NwSc}
\end{figure*}

From all the above analysis, the optimized sizes of nanowires (or insulators) and SCs in any heterostructures for observing MFs are proposed. Tab. \ref{tab1} presents the optimal sizes for various widely utilized heterostructures. In fact, the sizes of these heterostructures in current experiments have not achieved optimal values, which may be the main reason why MFs have not been observed. In the following, we will apply the developed theory to analyze the heterostructures of InSb-Al and InAs-Al, and provide specific suggestions for observing MFs.

\subsection*{Theoretical analysis for observing MFs}

For the heterostructures of InSb-Al or InAs-Al, the critical magnetic field and Lande factors are set as $B_{c}\simeq2\mathrm{T}$,
$g_{\mathrm{InAs}}=-15$, $g_{\mathrm{InSb}}=-40$ and $g_{s}=2$ \citep{Marcus_Effective_g_2018,Antipov_2018,Karsten_2018,Cao_2022}. Using these parameters, the maximum chemical potential window is obtained as 1.73\,meV (for InAs-Al) or 4.63\,meV (for InSb-Al). From this, the critical size of InAs (or InSb) nanowires is determined as $L_c=97$\,nm (or 74\,nm). Therefore, the optimal transverse sizes of the nanowire should be such that its width is less than 97\,nm (or 74\,nm), and its thickness ideally approaches zero \cite{sizes}. The currently fabricated InSb or InAs nanowires typically have transverse sizes of approximately 50–100\,nm \cite{kouwenhoven2025}. These sizes have not yet reached optimal sizes, which results in subband degeneracy. This degeneracy, caused by the finite-size effect of the nanowires, has also been observed in some hybrid nanowire systems \cite{Kammhuber_2016,Zhang_2022,Zhang_2024}. Consequently, multiple MFs, localized at each end, may emerge, as discussed in the previous sections. For example, in an InSb nanowire with $L_{w}^{y}=L_{w}^{z}=70\,\mathrm{nm}$, when the chemical potential lies within two degenerate subbands $(l_{w}^{y},l_{w}^{y})=(1,2)$ and $(2,1)$, two (though occasionally one \cite{Note1}) MFs are localized at each end within a specific  $\mu_w -B$ parameter region, as shown Fig. \ref{fig:Finite_size_NwSc}(a). In this scenario, for a given chemical potential, as the magnetic field increases, the ZBP transitions from zero to $2e^2/h$, and then to $4e^2/h$, as the number of MFs varies across different parameter regions. Thus, fabricating nanowires with optimal sizes is essential to ensure the realization of exactly one localized MF at each end.

For Al superconducting thin film used in experiments, its thickness is approximately $\sim10$\,nm \cite{Karsten_2018,Antipov_2018,Reeg_2018_Metallization_nw}. Around this thickness, both the induced shift in the chemical potential and the induced gap exhibit significant oscillations, as shown in Fig.
\ref{fig:Finite_size_NwSc}(d,\,e). This results in a periodic variation in the number of MFs as the thickness increases [see Fig. \ref{fig:Finite_size_NwSc}(b,c)]. This periodic structure with respect to the thickness can be experimentally verified by observing  the corresponding periodic variations of the ZBP between 0 and \(2e^2/h\). Since precisely controlling the thickness of the SC within half the Fermi wavelength $\sim 1$nm is unfeasible, the magnitude of the shift in the chemical potential and the induced gap are completely uncertain. This uncertainty will significantly reduce the likelihood of observing MFs. As shown in the blue region of Fig. \ref{fig:Finite_size_NwSc}(b) and \ref{fig:Finite_size_NwSc}(c), for a given magnetic field or chemical potential, MFs are absent over a range of thickness values. This occurs because, at certain thicknesses, the induced gap is too large or the nanowire's chemical potential falls outside its window. Additionally, thicknesses that result in a very small induced gap are not viable. This is because a small gap fails to protect the MFs from thermal excitations, impurities, and other perturbations. Therefore, further determining the proper induced energy gap and chemical potential window is more conducive to observing MFs. As shown in Fig. \ref{fig:Finite_size_NwSc}(f,\,g), the relationship between the chemical potential window and the induced energy gap indicates that for InAs-Al (or InSb-Al), the induced gap must satisfy  $\Delta_{\mathrm{ind}}<0.48(0.67)\Delta_{s}$, which corresponds to the correction factor $\lambda <1(2)$. Therefore, the strong proximity-tunneling effect significantly reduces the chemical potential window, making it impossible to observe MFs. More details can be found in the Method section. Based on the theoretical analysis above, when the control precision of the superconducting thickness does not reach the order of $1\text{\AA}$, i.e., it is much smaller than the Fermi wavelength of the SC, the optimal superconducting thickness should be on the order of the superconducting coherence length. Namely, for InAs-Al and InSb-Al, the thickness of Al thin films should be around 1\,$\mu\mathrm{m}$. However, the adopted thin films in current experiments are far smaller than such a thickness, making it difficult to observe MFs.

As an illustration, we have elucidated the effects of finite size on the observation of MFs in heterostructures of InSb and InAs, and futher determined their optimal sizes. For other superconducting heterostructures, size effects can also be analyzed in the same manner.

\section*{Discussions}\label{sec3}

We have analytically defined the boundary of the topological phase for finite-sized nanowire-superconductor heterostructures and determined the number of localized Majorana fermions (MFs) in the various phases. We have studied the finite-size effects of both the nanowire and the superconductor and identified their optimal sizes to ensure that only one MF exists at each end of the heterostructure. These findings are valid for any superconducting heterostructures, including nanowire-superconductor, insulator-superconductor heterostructures \cite{Yue_2023, yue2025}, and two quantum dot-superconductor systems \cite{Dvir2023,zhang2025robustzeromodespbtepb}. Specifically, the optimal sizes for the nanowire-superconductor and insulator-superconductor heterostructures have been outlined. Notably, the optimal thickness of the superconductor (Al films, $\sim$1000\,nm) is two orders of magnitude larger than the thickness of Al films currently utilized in experimental devices (e.g., InSb-Al and InAs-Al heterostructures). Our findings not only clarify why observing MFs is challenging in current experiments but also offer guidance for the sizes selection of heterostructure in future studies. 

In this work, we have studied the finite-size effects of both nanowire and superconductor in a heterostructure, where they exhibit rectangular transverse geometries \cite{Antipov_2018,Karsten_2018}. 
Furthermore, we have focused on investigating how the degeneracy effects of the nanowire's low-energy subbands, resulting from its finite size, influence the observed signals of MFs. In fact, for more geometric configurations adopted in experiments, such as circular
\cite{Das2012,Zhang2018,Vaiti_2020,Marco_2021_Nontopological_zero-bias,Zhang_2022},
hexagonal \cite{Levajac2023,Nick_2023}, or even irregular shapes, the degeneracy effect of subbands (including high-energy subbands) on observing MFs becomes more complex and warrants further investigation. In the derivations above, we have adopted the local tunneling approximation; that is, the tunneling processes with different momenta in the $z$-direction between the nanowire and superconductor do not have any interferences or correlations with each other \cite{Sau_Robustness_MF_2010,Qiao_2022}. When the weak non-local tunneling between different subbands of the nanowire is included, multiple MFs localized at the same end, originating from subband degeneracies, are likely to couple with one another. These coupled MFs give rise to the so-called quasi-Majorana fermions \cite{Wimmer_2019,Prada2020,Zhang_2022}. This phenomenon, inherently caused by the finite-size effect, is worth further attention.

We finally stress that even under ideal conditions (disregarding impurities and chemical potential inhomogeneities \cite{Sankar_Das_Sarma2023,kouwenhoven2025}), the narrow chemical potential window \cite{Qiao_2024}, and the unsuitable sizes of nanowires and superconductors (resulting in the degeneracy of energy levels, as well as oscillations in the chemical potential and the effecive gap) still present significant challenges in observing stable and clear signatures of MFs.

\backmatter

\bibliographystyle{sn-aps}
\bibliography{MFFSNc_v8}

\begin{thebibliography}{10}
\providecommand{\url}[1]{{#1}}
\providecommand{\urlprefix}{URL }
\providecommand{\doi}[1]{\url{https://doi.org/#1}}


\bibitem{Mourik_2012}
V.~Mourik, K.~Zuo, S.M. Frolov, S.R. Plissard, E.P.A.M. Bakkers, L.P.
  Kouwenhoven, Signatures of {M}ajorana fermions in {H}ybrid
  {S}uperconductor-{S}emiconductor {N}anowire {D}evices.
\newblock Science \textbf{336}(6084), 1003--1007 (2012).
\newblock \doi{10.1126/science.1222360}

\bibitem{Deng_2012}
M.T. Deng, C.L. Yu, G.Y. Huang, M.~Larsson, P.~Caroff, H.Q. Xu, Anomalous
  zero-bias conductance peak in a {N}b–{I}n{S}b nanowire–{N}b hybrid
  device.
\newblock Nano Letters \textbf{12}(12), 6414--6419 (2012).
\newblock \doi{10.1021/nl303758w}

\bibitem{Das2012}
A.~Das, Y.~Ronen, Y.~Most, Y.~Oreg, M.~Heiblum, H.~Shtrikman, Zero-bias peaks
  and splitting in an {Al–InAs} nanowire topological superconductor as a
  signature of majorana fermions.
\newblock Nature Physics \textbf{8}, 887--895 (2012).
\newblock \doi{10.1038/nphys2479}

\bibitem{Nichele_2017}
F.~Nichele, A.C.C. Drachmann, A.M. Whiticar, E.C.T. O'Farrell, H.J. Suominen,
  A.~Fornieri, T.~Wang, G.C. Gardner, C.~Thomas, A.T. Hatke, P.~Krogstrup, M.J.
  Manfra, K.~Flensberg, C.M. Marcus, Scaling of majorana zero-bias conductance
  peaks.
\newblock Phys. Rev. Lett. \textbf{119}, 136803 (2017).
\newblock \doi{10.1103/PhysRevLett.119.136803}

\bibitem{Gazibegovic2017}
S.~Gazibegovic, D.~Car, H.~Zhang, S.C. Balk, J.A. Logan, M.W.A. de~Moor, M.C.
  Cassidy, R.~Schmits, D.~Xu, G.~Wang, P.~Krogstrup, R.L.M.O. het Veld, K.~Zuo,
  Y.~Vos, J.~Shen, D.~Bouman, B.~Shojaei, D.~Pennachio, J.S. Lee, P.J. van
  Veldhoven, S.~Koelling, M.A. Verheijen, L.P. Kouwenhoven, C.J. Palmstrøm,
  E.P.A.M. Bakkers, Retracted article: Epitaxy of advanced nanowire quantum
  devices.
\newblock Nature \textbf{548}, 434--438 (2017).
\newblock \doi{10.1038/nature23468}

\bibitem{Zhang2018}
H.~Zhang, C.X. Liu, S.~Gazibegovic, D.~Xu, J.A. Logan, G.~Wang, N.~van Loo,
  J.D.S. Bommer, M.W.A. de~Moor, D.~Car, R.L.M.O. het Veld, P.J. van Veldhoven,
  S.~Koelling, M.A. Verheijen, M.~Pendharkar, D.J. Pennachio, B.~Shojaei, J.S.
  Lee, C.J. Palmstrøm, E.P.A.M. Bakkers, S.D. Sarma, L.P. Kouwenhoven,
  Retracted article: Quantized majorana conductance.
\newblock Nature \textbf{556}, 74--79 (2018).
\newblock \doi{10.1038/nature26142}

\bibitem{Marco_2021_Nontopological_zero-bias}
M.~Valentini, F.~Pe{\~{n}}aranda, A.~Hofmann, M.~Brauns, R.~Hauschild,
  P.~Krogstrup, P.~San-Jose, E.~Prada, R.~Aguado, G.~Katsaros, Nontopological
  zero-bias peaks in full-shell nanowires induced by flux-tunable andreev
  states.
\newblock Science \textbf{373}(6550), 82--88 (2021).
\newblock \doi{10.1126/science.abf1513}

\bibitem{Zhang_2022}
Z.~Wang, H.~Song, D.~Pan, Z.~Zhang, W.~Miao, R.~Li, Z.~Cao, G.~Zhang, L.~Liu,
  L.~Wen, R.~Zhuo, D.E. Liu, K.~He, R.~Shang, J.~Zhao, H.~Zhang, Plateau
  regions for zero-bias peaks within 5\% of the quantized conductance value
  $2{e}^{2}/h$.
\newblock Phys. Rev. Lett. \textbf{129}, 167702 (2022).
\newblock \doi{10.1103/PhysRevLett.129.167702}

\bibitem{Vaiti_2020}
S.~Vaitiekenas, G.W. Winkler, B.~van Heck, T.~Karzig, M.T. Deng, K.~Flensberg,
  L.I. Glazman, C.~Nayak, P.~Krogstrup, R.M. Lutchyn, C.M. Marcus, Flux-induced
  topological superconductivity in full-shell nanowires.
\newblock Science \textbf{367}(6485), eaav3392 (2020).
\newblock \doi{10.1126/science.aav3392}

\bibitem{Levajac2023}
V.~Levajac, J.Y. Wang, C.~Sfiligoj, M.~Lemang, J.C. Wolff, A.~Bordin,
  G.~Badawy, S.~Gazibegovic, E.P.A.M. Bakkers, L.P. Kouwenhoven, Subgap
  spectroscopy along hybrid nanowires by nm-thick tunnel barriers.
\newblock Nature Communications \textbf{14}, 6647 (2023).
\newblock \doi{10.1038/s41467-023-42422-z}

\bibitem{Dvir2023}
T.~Dvir, G.~Wang, N.~van Loo, C.X. Liu, G.P. Mazur, A.~Bordin, S.L.D. ten Haaf,
  J.Y. Wang, D.~van Driel, F.~Zatelli, X.~Li, F.K. Malinowski, S.~Gazibegovic,
  G.~Badawy, E.P.A.M. Bakkers, M.~Wimmer, L.P. Kouwenhoven, Realization of a
  minimal kitaev chain in coupled quantum dots.
\newblock Nature \textbf{614}, 445--450 (2023).
\newblock \doi{10.1038/s41586-022-05585-1}

\bibitem{Aghaee2025}
M.~Aghaee, A.A. Ramirez, Z.~Alam, R.~Ali, M.~Andrzejczuk, A.~Antipov,
  M.~Astafev, A.~Barzegar, B.~Bauer, J.~Becker, U.K. Bhaskar, A.~Bocharov,
  S.~Boddapati, D.~Bohn, J.~Bommer, L.~Bourdet, A.~Bousquet, S.~Boutin,
  L.~Casparis, B.J. Chapman, S.~Chatoor, A.W. Christensen, C.~Chua, P.~Codd,
  W.~Cole, P.~Cooper, F.~Corsetti, A.~Cui, P.~Dalpasso, J.P. Dehollain,
  G.~de~Lange, M.~de~Moor, A.~Ekefjärd, T.E. Dandachi, J.C.E. Saldaña,
  S.~Fallahi, L.~Galletti, G.~Gardner, D.~Govender, F.~Griggio, R.~Grigoryan,
  S.~Grijalva, S.~Gronin, J.~Gukelberger, M.~Hamdast, F.~Hamze, E.B. Hansen,
  S.~Heedt, Z.~Heidarnia, J.H. Zamorano, S.~Ho, L.~Holgaard, J.~Hornibrook,
  J.~Indrapiromkul, H.~Ingerslev, L.~Ivancevic, T.~Jensen, J.~Jhoja, J.~Jones,
  K.V. Kalashnikov, R.~Kallaher, R.~Kalra, F.~Karimi, T.~Karzig, E.~King, M.E.
  Kloster, C.~Knapp, D.~Kocon, J.V. Koski, P.~Kostamo, M.~Kumar, T.~Laeven,
  T.~Larsen, J.~Lee, K.~Lee, G.~Leum, K.~Li, T.~Lindemann, M.~Looij, J.~Love,
  M.~Lucas, R.~Lutchyn, M.H. Madsen, N.~Madulid, A.~Malmros, M.~Manfra,
  D.~Mantri, S.B. Markussen, E.~Martinez, M.~Mattila, R.~McNeil, A.B. Mei, R.V.
  Mishmash, G.~Mohandas, C.~Mollgaard, T.~Morgan, G.~Moussa, C.~Nayak, J.H.
  Nielsen, J.M. Nielsen, W.H.P. Nielsen, B.~Nijholt, M.~Nystrom,
  E.~O’Farrell, T.~Ohki, K.~Otani, B.P. Wütz, S.~Pauka, K.~Petersson,
  L.~Petit, D.~Pikulin, G.~Prawiroatmodjo, F.~Preiss, E.P. Morejon,
  M.~Rajpalke, C.~Ranta, K.~Rasmussen, D.~Razmadze, O.~Reentila, D.J. Reilly,
  Y.~Ren, K.~Reneris, R.~Rouse, I.~Sadovskyy, L.~Sainiemi, I.~Sanlorenzo,
  E.~Schmidgall, C.~Sfiligoj, M.B. Shah, K.~Simoes, S.~Singh, S.~Sinha,
  T.~Soerensen, P.~Sohr, T.~Stankevic, L.~Stek, E.~Stuppard, H.~Suominen,
  J.~Suter, S.~Teicher, N.~Thiyagarajah, R.~Tholapi, M.~Thomas, E.~Toomey,
  J.~Tracy, M.~Turley, S.~Upadhyay, I.~Urban, K.V. Hoogdalem, D.J.V. Woerkom,
  D.V. Viazmitinov, D.~Vogel, J.~Watson, A.~Webster, J.~Weston, G.W. Winkler,
  D.~Xu, C.K. Yang, E.~Yucelen, R.~Zeisel, G.~Zheng, J.~Zilke, M.A. Quantum,
  Interferometric single-shot parity measurement in {InAs–Al} hybrid devices.
\newblock Nature \textbf{638}, 651--655 (2025).
\newblock \doi{10.1038/s41586-024-08445-2}

\bibitem{zhang2025robustzeromodespbtepb}
S.~Zhang, W.~Song, Z.~Li, Z.~Yu, R.~Li, Y.~Wang, Z.~Yan, J.~Xu, Z.~Wang,
  Y.~Gao, S.~Yang, L.~Yang, X.~Feng, T.~Wang, Y.~Zang, L.~Li, R.~Shang, Q.K.
  Xue, K.~He, H.~Zhang.
\newblock Robust zero modes in {PbTe-Pb} hybrid nanowires (2025)

\bibitem{A_Yu_Kitaev_2001}
A.Y. Kitaev, Unpaired majorana fermions in quantum wires.
\newblock Physics-Uspekhi \textbf{44}(10S), 131 (2001).
\newblock \doi{10.1070/1063-7869/44/10S/S29}

\bibitem{Kitaev_2003}
A.~Kitaev, Fault-tolerant quantum computation by anyons.
\newblock Annals of Physics \textbf{303}(1), 2--30 (2003).
\newblock \doi{https://doi.org/10.1016/S0003-4916(02)00018-0}

\bibitem{Nayak_2008}
C.~Nayak, S.H. Simon, A.~Stern, M.~Freedman, S.~Das~Sarma, Non-abelian anyons
  and topological quantum computation.
\newblock Rev. Mod. Phys. \textbf{80}, 1083--1159 (2008).
\newblock \doi{10.1103/RevModPhys.80.1083}

\bibitem{Oreg_2010}
Y.~Oreg, G.~Refael, F.~von Oppen, Helical liquids and majorana bound states in
  quantum wires.
\newblock Phys. Rev. Lett. \textbf{105}, 177002 (2010).
\newblock \doi{10.1103/PhysRevLett.105.177002}

\bibitem{Alicea_2012}
J.~Alicea, New directions in the pursuit of majorana fermions in solid state
  systems.
\newblock Reports on Progress in Physics \textbf{75}(7), 076501 (2012).
\newblock \doi{10.1088/0034-4885/75/7/076501}

\bibitem{Lutchyn2018}
R.M. Lutchyn, E.P.A.M. Bakkers, L.P. Kouwenhoven, P.~Krogstrup, C.M. Marcus,
  Y.~Oreg, Majorana zero modes in superconductor-semiconductor
  heterostructures.
\newblock Nature Reviews Materials \textbf{3}, 52--68 (2018).
\newblock \doi{10.1038/s41578-018-0003-1}

\bibitem{Prada2020}
E.~Prada, P.~San-Jose, M.W.A. de~Moor, A.~Geresdi, E.J.H. Lee, J.~Klinovaja,
  D.~Loss, J.~Nyg\r{a}rd, R.~Aguado, L.P. Kouwenhoven, From andreev to majorana
  bound states in hybrid superconductor–semiconductor nanowires.
\newblock Nature Reviews Physics \textbf{2}, 575--594 (2020).
\newblock \doi{10.1038/s42254-020-0228-y}

\bibitem{Qiao_2022}
G.J. Qiao, S.W. Li, C.P. Sun, Magnetic field constraint for majorana zero modes
  in a hybrid nanowire.
\newblock Phys. Rev. B \textbf{106}, 104517 (2022).
\newblock \doi{10.1103/PhysRevB.106.104517}

\bibitem{Qiao_2024}
G.J. Qiao, X.~Yue, C.P. Sun, Dressed majorana fermion in a hybrid nanowire.
\newblock Phys. Rev. Lett. \textbf{133}, 266605 (2024).
\newblock \doi{10.1103/PhysRevLett.133.266605}

\bibitem{Li_2014}
S.W. Li, Z.Z. Li, C.Y. Cai, C.P. Sun, Probing zero modes of a defect in a
  kitaev quantum wire.
\newblock Phys. Rev. B \textbf{89}, 134505 (2014).
\newblock \doi{10.1103/PhysRevB.89.134505}

\bibitem{Zhang_Fan_2023}
F.~Zhang, J.~Gu, H.T. Quan, Full counting statistics, fluctuation relations,
  and linear response properties in a one-dimensional kitaev chain.
\newblock Phys. Rev. E \textbf{108}, 024110 (2023).
\newblock \doi{10.1103/PhysRevE.108.024110}

\bibitem{Yue_2023}
X.~Yue, G.J. Qiao, C.P. Sun, Refined majorana phase diagram in a topological
  insulator--superconductor hybrid system.
\newblock Phys. Rev. B \textbf{108}, 195405 (2023).
\newblock \doi{10.1103/PhysRevB.108.195405}

\bibitem{Potter_2011}
A.C. Potter, P.A. Lee, Majorana end states in multiband microstructures with
  rashba spin-orbit coupling.
\newblock Phys. Rev. B \textbf{83}, 094525 (2011).
\newblock \doi{10.1103/PhysRevB.83.094525}

\bibitem{Lutchyn_2011_in_Multiband_Nanowires}
R.M. Lutchyn, T.D. Stanescu, S.~Das~Sarma, Search for majorana fermions in
  multiband semiconducting nanowires.
\newblock Phys. Rev. Lett. \textbf{106}, 127001 (2011).
\newblock \doi{10.1103/PhysRevLett.106.127001}

\bibitem{Stanescu_2011_MF_in_semiconductor_nanowires}
T.D. Stanescu, R.M. Lutchyn, S.~Das~Sarma, Majorana fermions in semiconductor
  nanowires.
\newblock Phys. Rev. B \textbf{84}, 144522 (2011).
\newblock \doi{10.1103/PhysRevB.84.144522}

\bibitem{Liu_Jie_2012}
J.~Liu, A.C. Potter, K.T. Law, P.A. Lee, Zero-bias peaks in the tunneling
  conductance of spin-orbit-coupled superconducting wires with and without
  majorana end-states.
\newblock Phys. Rev. Lett. \textbf{109}, 267002 (2012).
\newblock \doi{10.1103/PhysRevLett.109.267002}

\bibitem{Karsten_2018}
A.E.G. Mikkelsen, P.~Kotetes, P.~Krogstrup, K.~Flensberg, Hybridization at
  superconductor-semiconductor interfaces.
\newblock Phys. Rev. X \textbf{8}, 031040 (2018).
\newblock \doi{10.1103/PhysRevX.8.031040}

\bibitem{Antipov_2018}
A.E. Antipov, A.~Bargerbos, G.W. Winkler, B.~Bauer, E.~Rossi, R.M. Lutchyn,
  Effects of gate-induced electric fields on semiconductor majorana nanowires.
\newblock Phys. Rev. X \textbf{8}, 031041 (2018).
\newblock \doi{10.1103/PhysRevX.8.031041}

\bibitem{Wimmer_2019}
A.~Vuik, B.~Nijholt, A.R. Akhmerov, M.~Wimmer, Reproducing topological
  properties with quasi-majorana states.
\newblock SciPost Phys. \textbf{7}, 061 (2019).
\newblock \doi{10.21468/SciPostPhys.7.5.061}

\bibitem{Potter_2010}
A.C. Potter, P.A. Lee, Multichannel generalization of kitaev's majorana end
  states and a practical route to realize them in thin films.
\newblock Phys. Rev. Lett. \textbf{105}, 227003 (2010).
\newblock \doi{10.1103/PhysRevLett.105.227003}

\bibitem{Roy_2013}
D.~Roy, N.~Bondyopadhaya, S.~Tewari, Topologically trivial zero-bias
  conductance peak in semiconductor majorana wires from boundary effects.
\newblock Phys. Rev. B \textbf{88}, 020502 (2013).
\newblock \doi{10.1103/PhysRevB.88.020502}

\bibitem{Jelena_2017_Finite_size_effects}
C.~Reeg, D.~Loss, J.~Klinovaja, Finite-size effects in a nanowire strongly
  coupled to a thin superconducting shell.
\newblock Phys. Rev. B \textbf{96}, 125426 (2017).
\newblock \doi{10.1103/PhysRevB.96.125426}

\bibitem{Reeg_2018_Metallization_nw}
C.~Reeg, D.~Loss, J.~Klinovaja, Metallization of a rashba wire by a
  superconducting layer in the strong-proximity regime.
\newblock Phys. Rev. B \textbf{97}, 165425 (2018).
\newblock \doi{10.1103/PhysRevB.97.165425}

\bibitem{Legg_2022_Metallization_Ti_NW}
H.F. Legg, D.~Loss, J.~Klinovaja, Metallization and proximity superconductivity
  in topological insulator nanowires.
\newblock Phys. Rev. B \textbf{105}, 155413 (2022).
\newblock \doi{10.1103/PhysRevB.105.155413}

\bibitem{Winkler_2019}
G.W. Winkler, A.E. Antipov, B.~van Heck, A.A. Soluyanov, L.I. Glazman,
  M.~Wimmer, R.M. Lutchyn, Unified numerical approach to topological
  semiconductor-superconductor heterostructures.
\newblock Phys. Rev. B \textbf{99}, 245408 (2019).
\newblock \doi{10.1103/PhysRevB.99.245408}

\bibitem{ZZ2025}
Z.L. Zhang, G.J. Qiao, C.P. Sun.
\newblock Poor man's majoranon in two quantum dots dressed by superconducting
  quasi-excitations (2025).
\newblock \urlprefix\url{https://arxiv.org/abs/2506.10367}

\bibitem{yue2025}
X.~Yue, G.J. Qiao, C.P. Sun.
\newblock Finite thickness effects on metallization vs. chiral majorana
  fermions (2025).
\newblock \urlprefix\url{https://arxiv.org/abs/2506.16093}

\bibitem{Xue2004}
Y.~Guo, Y.F. Zhang, X.Y. Bao, T.Z. Han, Z.~Tang, L.X. Zhang, W.G. Zhu, E.G.
  Wang, Q.~Niu, Z.Q. Qiu, J.F. Jia, Z.X. Zhao, Q.K. Xue, Superconductivity
  modulated by quantum size effects.
\newblock Science \textbf{306}(5703), 1915--1917 (2004).
\newblock \doi{10.1126/science.1105130}

\bibitem{Tewari_2012}
S.~Tewari, J.D. Sau, Topological invariants for spin-orbit coupled
  superconductor nanowires.
\newblock Phys. Rev. Lett. \textbf{109}, 150408 (2012).
\newblock \doi{10.1103/PhysRevLett.109.150408}

\bibitem{Budich_2013}
J.C. Budich, E.~Ardonne, Equivalent topological invariants for one-dimensional
  majorana wires in symmetry class $d$.
\newblock Phys. Rev. B \textbf{88}, 075419 (2013).
\newblock \doi{10.1103/PhysRevB.88.075419}

\bibitem{Chiu_2016_Classification_topological_symmetries}
C.K. Chiu, J.C.Y. Teo, A.P. Schnyder, S.~Ryu, Classification of topological
  quantum matter with symmetries.
\newblock Rev. Mod. Phys. \textbf{88}, 035005 (2016).
\newblock \doi{10.1103/RevModPhys.88.035005}

\bibitem{Clogston_1962}
A.M. Clogston, Upper limit for the critical field in hard superconductors.
\newblock Phys. Rev. Lett. \textbf{9}, 266--267 (1962).
\newblock \doi{10.1103/PhysRevLett.9.266}

\bibitem{Maki1964}
K.~Maki, T.~Tsuneto, Pauli paramagnetism and superconducting state.
\newblock Progress of Theoretical Physics \textbf{31}, 945--956 (1964).
\newblock \doi{10.1143/PTP.31.945}

\bibitem{Reeg_transport_Signature_2017}
C.~Reeg, D.L. Maslov, Transport signatures of topological superconductivity in
  a proximity-coupled nanowire.
\newblock Phys. Rev. B \textbf{95}, 205439 (2017).
\newblock \doi{10.1103/PhysRevB.95.205439}

\bibitem{Marcus_Effective_g_2018}
S.~Vaitiek\ifmmode~\dot{e}\else \.{e}\fi{}nas, M.T. Deng, J.~Nyg\r{a}rd,
  P.~Krogstrup, C.M. Marcus, Effective $g$ factor of subgap states in hybrid
  nanowires.
\newblock Phys. Rev. Lett. \textbf{121}, 037703 (2018).
\newblock \doi{10.1103/PhysRevLett.121.037703}

\bibitem{Sau_Robustness_MF_2010}
J.D. Sau, R.M. Lutchyn, S.~Tewari, S.~Das~Sarma, Robustness of majorana
  fermions in proximity-induced superconductors.
\newblock Phys. Rev. B \textbf{82}, 094522 (2010).
\newblock \doi{10.1103/PhysRevB.82.094522}

\bibitem{Parameterspace}
The parameter space should includes all parameters of the nanowire, {SC}, and
  their tunneling strength. {S}ince most parameters are determined by the
  chosen materials, we primarily focus on the ones that can be experimentally
  adjusted.

\bibitem{Wimmer_2011}
M.~Wimmer, A.R. Akhmerov, J.P. Dahlhaus, C.W.J. Beenakker, Quantum point
  contact as a probe of a topological superconductor.
\newblock New Journal of Physics \textbf{13}(5), 053016 (2011).
\newblock \doi{10.1088/1367-2630/13/5/053016}

\bibitem{Nick_2023}
N.~van Loo, G.P. Mazur, T.~Dvir, G.~Wang, R.C. Dekker, J.Y. Wang, M.~Lemang,
  C.~Sfiligoj, A.~Bordin, D.~van Driel, G.~Badawy, S.~Gazibegovic, E.P.A.M.
  Bakkers, L.P. Kouwenhoven, Electrostatic control of the proximity effect in
  the bulk of semiconductor-superconductor hybrids.
\newblock Nature Communications \textbf{14}, 3325 (2023).
\newblock \doi{10.1038/s41467-023-39044-w}

\bibitem{Cao_2022}
Z.~Cao, D.E. Liu, W.X. He, X.~Liu, K.~He, H.~Zhang, Numerical study of
  {PbTe-Pb} hybrid nanowires for engineering majorana zero modes.
\newblock Phys. Rev. B \textbf{105}, 085424 (2022).
\newblock \doi{10.1103/PhysRevB.105.085424}

\bibitem{sizes}
The smaller transverse dimension of a real nanowire material should be taken as
  $\sim$1nm. {T}he corresponding energy level spacing reaches approximately
  1e{V}, which effectively prevents subband degeneracy in the nanowire.

\bibitem{kouwenhoven2025}
L.~Kouwenhoven, Perspective on {M}ajorana bound-states in hybrid
  superconductor-semiconductor nanowires.
\newblock Mod. Phys. Lett. B \textbf{39}(03), 2540002 (2025).
\newblock \doi{10.1142/S0217984925400020}

\bibitem{Kammhuber_2016}
J.~Kammhuber, M.C. Cassidy, H.~Zhang, {\"O}.~G{\"u}l, F.~Pei, M.W.A. de~Moor,
  B.~Nijholt, K.~Watanabe, T.~Taniguchi, D.~Car, S.R. Plissard, E.P.A.M.
  Bakkers, L.P. Kouwenhoven, Conductance quantization at zero magnetic field in
  {InSb} nanowires.
\newblock Nano Letters \textbf{16}, 3482--3486 (2016).
\newblock \doi{10.1021/acs.nanolett.6b00051}

\bibitem{Zhang_2024}
Y.~Wang, W.~Song, Z.~Cao, Z.~Yu, S.~Yang, Z.~Li, Y.~Gao, R.~Li, F.~Chen,
  Z.~Geng, L.~Yang, J.~Xu, Z.~Wang, S.~Zhang, X.~Feng, T.~Wang, Y.~Zang, L.~Li,
  R.~Shang, Q.K. Xue, D.E. Liu, K.~He, H.~Zhang, Gate-tunable subband
  degeneracy in semiconductor nanowires.
\newblock Proceedings of the National Academy of Sciences \textbf{121}(27),
  e2406884121 (2024).
\newblock \doi{10.1073/pnas.2406884121}

\bibitem{Note1}
For a given magnetic field, one {M}ajorana fermion at each end can be observed
  within a narrow chemical potential window of approximately $2\,
  \mu\text{eV}$.

\bibitem{Sankar_Das_Sarma2023}
S.D. Sarma, In search of {M}ajorana.
\newblock Nature Physics \textbf{19}, 165--170 (2023).
\newblock \doi{10.1038/s41567-022-01900-9}

\bibitem{Cole_2016}
W.S. Cole, J.D. Sau, S.~Das~Sarma, Proximity effect and majorana bound states
  in clean semiconductor nanowires coupled to disordered superconductors.
\newblock Phys. Rev. B \textbf{94}, 140505 (2016).
\newblock \doi{10.1103/PhysRevB.94.140505}

\bibitem{2010Proximity}
T.D. Stanescu, J.D. Sau, R.M. Lutchyn, S.~Das~Sarma, Proximity effect at the
  superconductor--topological insulator interface.
\newblock Phys. Rev. B \textbf{81}, 241310 (2010).
\newblock \doi{10.1103/PhysRevB.81.241310}

\bibitem{Stanescu_2017_Proximity-induced_low-energy_renormalization}
T.D. Stanescu, S.~Das~Sarma, Proximity-induced low-energy renormalization in
  hybrid semiconductor-superconductor majorana structures.
\newblock Phys. Rev. B \textbf{96}, 014510 (2017).
\newblock \doi{10.1103/PhysRevB.96.014510}

\bibitem{Liu_2017_Andreev_bound_states}
C.X. Liu, J.D. Sau, T.D. Stanescu, S.~Das~Sarma, Andreev bound states versus
  majorana bound states in quantum dot-nanowire-superconductor hybrid
  structures: Trivial versus topological zero-bias conductance peaks.
\newblock Phys. Rev. B \textbf{96}, 075161 (2017).
\newblock \doi{10.1103/PhysRevB.96.075161}

\end{thebibliography}
\clearpage
\section*{Methods}\label{sec4}
\subsection*{Finite-size Model for Majorana}
We characterize the finite-size nanowire-superconductor heterostructure by the Slab model
\cite{Antipov_2018,Karsten_2018}. The Hamiltonian of this finite-size system is $H=H_{w}+H_{s}+H_{t}$,
where the nanowire is described by \cite{Potter_2011,Lutchyn_2011_in_Multiband_Nanowires,Stanescu_2011_MF_in_semiconductor_nanowires,Liu_Jie_2012}
\begin{align}
	H_{w}&=\sum_{\boldsymbol{n}_{w}}d_{\boldsymbol{n}_{w}}^{\dagger}(h_{w}\sigma_{x}-\tilde{\mu}_{w})d_{\boldsymbol{n}_{w}}-\frac{t_{w}}{2}(d_{\boldsymbol{n}_{w}}^{\dagger}d_{\boldsymbol{n}_{w}+\boldsymbol{\delta}_{w}}+\mathrm{H.c})\nonumber \\
	&-\frac{\mathrm{i}}{2}\sum_{\boldsymbol{n}_{w}}\sum_{i}d_{\boldsymbol{n}_{w}}^{\dagger}\alpha_{i}(\boldsymbol{\sigma}\times\hat{\boldsymbol{e}}_{z})_{i}d_{\boldsymbol{n}_{w}+\boldsymbol{e}_{i}}+\mathrm{H.c.}
	\label{eq:Hw}
\end{align}
A finite-size $s$-wave SC or the superconducting shell is in contact with the finite-size nanowire, and such an SC is described by the BCS Hamiltonian \cite{Jelena_2017_Finite_size_effects,Stanescu_2011_MF_in_semiconductor_nanowires,Reeg_2018_Metallization_nw}
\begin{align}
	H_{s}&=\sum_{\boldsymbol{n}_{s}}c_{\boldsymbol{n}_{s}}^{\dagger}(h_{s}\sigma_{x}-\tilde{\mu}_{s})c_{\boldsymbol{n}_{s}}-\frac{t_{s}}{2}(c_{\boldsymbol{n}_{s}}^{\dagger}c_{\boldsymbol{n}_{s}+\boldsymbol{\delta}_{s}}+\mathrm{H.c})\nonumber \\
	&+\sum_{\boldsymbol{n}_{s}}\Delta_{s}c_{\boldsymbol{n}_{s}\uparrow}^{\dagger}c_{\boldsymbol{n}_{s}\downarrow}^{\dagger}+\mathrm{H.c.}
	\label{eq:Hs}
\end{align}
And the proximity tunneling between the nanowire and SC is \cite{Stanescu_2011_MF_in_semiconductor_nanowires,Cole_2016,Reeg_2018_Metallization_nw}.
\begin{align}
	H_{t}=\sum_{n_{x},n_{y}}T(c_{n_{x},n_{y}^{s},N_{z}^{s}}^{\dagger}d_{n_{x},n_{y}^{w},1}+\mathrm{H.c.}).
\end{align}  
Here, $\boldsymbol{n}_{\alpha}=(n_{x},n_{y}^{\alpha},n_{z}^{\alpha})$ labels lattice sites, where $\alpha\in\{w,s\}$ denotes the nanowire and SC, respectively. The notation $N_{i}^{\alpha}$ with $i\in\{x,y,z\}$
represents the total number of sites along the $i$-direction, and
$\boldsymbol{\delta}\in\{\boldsymbol{e}_{x},\boldsymbol{e}_{y},\boldsymbol{e}_{z}\}$
is a unit vector connecting nearest-neighboring sites. $b_{\boldsymbol{n}_{\alpha}}=[b_{\boldsymbol{n}_{\alpha}\uparrow},b_{\boldsymbol{n}_{\alpha}\downarrow}]^{T}$
with $b(=d,c)$ is defined by an annihilation operator of the electron
in site-$n_{\alpha}$ and $\sigma_{x,y,z}$ are the Pauli matrices.
The effective chemical potential $\tilde{\mu}_{\alpha}=\mu_{\alpha}-(1+\cos[\pi/(N_{z}^{\alpha}+1)])t_{\alpha}+\tilde{t}_{\alpha}^{y} \cos[\pi/(N_{y}^{\alpha}+1)])$,
with $\tilde{t}_{w}^{y}=\sqrt{t_{w}^{2}+\alpha_{y}^{2}}$ and $\tilde{t}_{s}^{y}=t_{s}$,
determines the level reference. $\mu_{\alpha}$ is the chemical potential, $t_{\alpha}$ is the hopping strength, and $h_{\alpha}$ is the Zeeman energy \cite{Clogston_1962,Maki1964,Reeg_transport_Signature_2017,Marcus_Effective_g_2018}. $\alpha_{x}$ and $\alpha_{y}$ is Rashba spin-orbit coupling in $x-y$ plane, and $\Delta_{s}$ is the superconducting pairing strength. 

\subsection*{Topological invariant and number of MFs}
We take periodic boundary conditions along the $x$-direction and do the Fourier transform on the operators $b_{n_{\alpha},\sigma}$. Then the Hamiltonian is rewritten as $H=(1/2)\sum_{k_{x}}\mathbf{C}_{k_{x}}^{\dagger}\cdot\mathbf{H}(k_{x})\cdot\mathbf{C_{\mathrm{\mathit{k_{x}}}}}$, where $\mathbf{C}_{k_{x}}=[\mathbf{C}_{k_{x}}^{e},\mathbf{C}_{k_{x}}^{h}]$ with $\mathbf{C}_{k_{x}}^{e}=[\mathrm{C}_{k_{x}}^{e,w},\mathrm{C}_{k_{x}}^{e,s}]^{T}$ and $\mathbf{C}_{k_{x}}^{h}=\left(\mathbf{C}_{-k_{x}}^{e}\right)^{\dagger}$. Here, $\mathrm{C}_{k_{x}}^{e,\alpha}$ and $\mathrm{C}_{k_{x}}^{h,\alpha}$ describe the annihilation operators for the electron and hole, respectively. Using the obtained Hamiltonian matrix $\mathbf{H}(k_x)$, we calculate the topological invariant as follows:
\begin{align}
	M=\mathrm{sgn}\{\mathrm{Pf}[\mathbf{H}_{\gamma}(0)]\mathrm{Pf}[\mathbf{H}_{\gamma}(\pi)]\},
	\label{IR}
\end{align}
where $\mathbf{H}_{\gamma}(k_x)=-i\mathbf{U}\cdot \mathbf{H}(k_{x})\cdot \mathbf{U}^{\dagger}$ with the unitary matrix
\begin{align} 
	\mathbf{U}=\frac{1}{\sqrt{2}}\begin{bmatrix}\mathbb{I} & \mathbb{I}\\
		-i\mathbb{I} & i\mathbb{I}
	\end{bmatrix}.
\end{align}
By Eq. \eqref{IR}, we obtain the analytical result for the topological invariant in Eq. \eqref{eq:TIn}. Thus, $M=-1$ defines the parameter region corresponding to the topological phase.

Then we determine the number of MFs in different phase regions. Under open boundary conditions along the $x$-direction, we diagonalize the Hamiltonian as $H=\sum_{\nu}E_{\nu}\eta_{\nu}^{\dagger}\eta_{\nu}$, where $\eta_{\nu}$ is the annihilation operator of the quasiparticle of the total hybrid system. By the definition of dressed MFs: $\eta_{\nu}^{\dagger}=\eta_{\nu}$ \cite{Qiao_2024},  we obtain that, for any subband of the
nanowire denoted by the transverse mode $\boldsymbol{k}_{w}^{\perp}$, the existence condition for MFs and their number can be determined.

Furthermore, to prevent the system from transitioning into a non-topological phase caused by extensive low-energy subband degeneracy in the nanowire, we outline the necessary constraints on the transverse sizes of the nanowire and further determine its optimal size to ensure that only one Majorana fermion exists at each end of the heterostructure.

\subsection*{Optimization of the induced gap}
In Eqs. (\ref{mu_shift}, \ref{induced_gap}), the shift of the chemical potential and the effective pairing strength have been obtained. Through the renormalization factor $Z=1/(\lambda+1)$, they directly connect the observed quantity--band shift $\mu_{\mathrm{ind}}\simeq Z\mu_{\mathrm{shift}}$ and the induced gap $\Delta_{\mathrm{ind}}\simeq Z\Delta_{\mathrm{eff}}$ \citep{yue2025,2010Proximity,Stanescu_2017_Proximity-induced_low-energy_renormalization}. Since the induced gap can be used to estimate the magnitude of the energy gap, while the window of chemical potential determine the range of chemical potential for observing MF, the larger they are, the more favorable it is for observing MFs. Given the Zeeman energy $h_{\alpha}=\mu_{B}g_{\alpha}B/2$ and the dependence of the
superconducting gap on the magnetic field $\Delta_{s}(B)=\Delta_{s}\sqrt{1-B^{2}/B_{c}^{2}}$
\citep{Liu_2017_Andreev_bound_states}, we can optimize $\mu_{c}$
and $\Delta_{\mathrm{ind}}$ with respect to magnetic field $B$ and
the correction factor $\lambda$. Here other parameters, including the
Lande factors $g_{\alpha}$, the Bohr magneton $\mu_{B}$, and the
critical magnetic field $B_{c}$ are determined by the specific material of nanowire and SC. Specifically, by following the functional relationships 
\begin{equation}
\begin{aligned}
\mu_{c}(\lambda,B)&=2\sqrt{\mu_{B}^2 [g_{w}+\lambda g_{s}]^{2} B^2-(\lambda \Delta_{s})^{2}},\\
\Delta_{\mathrm{ind}}(\lambda,B)&= Z\Delta_{\mathrm{eff}}=\frac{\lambda}{\lambda+1} \sqrt{1-\frac{B^2}{B^{2}_{c}}} \Delta_s,
\end{aligned}
\end{equation}
where $\lambda$ and $B$ are variables, we obtain the constraint relation between the chemical potential window and the induced energy gap, as shown in Fig. \ref{fig:Finite_size_NwSc}(f) and (g). At the boundary, the chemical potential window and the induced energy gap exhibit a trade-off relationship: as the chemical potential window increases, the induced energy gap typically decreases, and vice versa. In the limiting cases, the chemical potential window
reaches its maximum $1.73$meV (for InAs-Al) or 4.63meV (for InSb-Al) when the induced energy gap vanishes. Conversely, the induced energy gap is maximized at 0.48$\Delta_{s}$ or 0.67$\Delta_{s}$ when the
chemical potential window is zero. For observing MF, the chemical potential window and energy gap should take moderate values, around 1meV (2.5meV) and 0.3$\Delta_{s}$ (0.5$\Delta_{s}$) for InAs-Al
(InSb-Al). Correspondingly, the magnetic field is approximately $\sim0.8\mathrm{T}$.

Additional details regarding all the above derivation processes can be found in the Supplementary Information (SI).

\section*{Acknowledgements}

The authors appreciate quite much for the helpful discussion with Sheng-Wen Li in BIT. This study is supported by the National Natural Science Foundation of China (NSFC) (Grant No. 12088101) and NSAF (Grant No. U2230401).

\section*{Author contributions}
C. P. Sun and G. J. Qiao conceived and initiated the project; C. P. Sun supervised the project. G. J. Qiao, Z. L. Zhang, and X. Yue performed the calculations. Z. L. Zhang and G. J. Qiao prepared the figures and Supplementary Information. All authors discussed the results and made substantial contributions to the writing of the manuscript.
\section*{Competing interests}
The authors declare no competing interests
\end{document}